\documentclass[aps,prl,twocolumn,superscriptaddress,amsfont,graphicx,nofootinbib,preprintnumbers]{revtex4}%
\usepackage{xcolor,graphicx,epsfig}
\usepackage{ifpdf}
\usepackage{amsmath}
\usepackage{bm}
\usepackage{color}
\usepackage[english]{babel}
\usepackage{graphicx}%
\usepackage{amsfonts}%
\usepackage{amssymb}
\usepackage{braket}
\usepackage{hyperref}

\definecolor{nicered}{rgb}{0.7,0.1,0.1}
\definecolor{nicegreen}{rgb}{0.1,0.5,0.1}
\hypersetup{colorlinks,citecolor= nicegreen,linkcolor= nicered}

\newcommand{\inv}[0]{\text{invisibles}}
\newcommand{\stat}[0]{\text{stat.}}
\newcommand{\sys}[0]{\text{sys.}}

\usepackage[normalem]{ulem}

\def\mysection#1{{{\bf #1}.~}}
\begin{document}

\title{Implications of an enhanced $\boldsymbol{B \to K \nu \bar \nu}$ branching ratio}
\author{Rigo Bause}
\email[Electronic address: ]{rigo.bause@tu-dortmund.de}
\affiliation{TU Dortmund University, Department of Physics, Otto-Hahn-Str.4, D-44221 Dortmund, Germany}
\author{Hector Gisbert}
\email[Electronic address: ]{hector.gisbert@pd.infn.it}
\affiliation{Istituto Nazionale di Fisica Nucleare (INFN), Sezione di Padova,,Via F. Marzolo 8, 35131 Padova, Italy}
\affiliation{Dipartimento di Fisica e Astronomia
``Galileo Galilei'', Universita di Padova,Via F. Marzolo 8, 35131 Padova, Italy}
\author{Gudrun Hiller}
\email[Electronic address: ]{ghiller@physik.uni-dortmund.de}
\affiliation{TU Dortmund University, Department of Physics, Otto-Hahn-Str.4, D-44221 Dortmund, Germany}
\affiliation{Department of Physics and Astronomy, University of Sussex, Brighton, BN1 9QH, U.K.}

\begin{abstract}
Rare decays mediated by $b \to s \nu \bar \nu$ transitions have been reported by the Belle II experiment.
The branching ratio of the decay $B^+ \to K^+ \nu \bar \nu$ is found to be enhanced with respect to the standard model value.
If taken at face value, the implications are profound:
either lepton flavor universality is violated at the (multi)-TeV-scale,  or light new physics is involved.
This holds in general if $\mathcal{B}(B^+ \to K^+ \nu \bar \nu)$ exceeds $1.2 \cdot 10^{-5} \, (1.3 \cdot 10^{-5})$ at $1 \sigma$ ($2 \sigma$), 
which tightens with a decreasing upper limit on $\mathcal{B}(B \to K^*\nu \bar \nu)$, that is in reach of the Belle II experiment.
In view of the strong constraints on electron-muon universality violation in $|\Delta b|=|\Delta s|=1$ processes,
viable explanations are heavy, $(5-10)$-TeV tree-level new physics mediators that couple only to tau-flavors,
or lepton flavor violating ones. In addition, couplings of similar size  to both left- and right-handed quarks are generically required,
implying non-minimal BSM sectors which are carefully balanced against flavor constraints. The decay $B_s^0 \to \inv$  can shed light on whether new physics is light or heavy. 
In the former case, branching ratios can be as large as 
$10^{-5}$.

\end{abstract}

\maketitle

\section{Introduction}

Flavor changing neutral currents (FCNCs) are ideally suited to test the standard model (SM) due to their suppression by loops, flavor mixing angles
and the Glashow-Iliopoulos-Maiani-mechanism. $|\Delta b|=|\Delta s|=1$ transitions have been targeted for searches for new physics since the end of the past millennium.
Rare $B$-meson decays into muons, and to a lesser degree electrons, or a photon have received particular scrutiny at the B-factories Belle and BaBar, and the LHC-experiments
\cite{ParticleDataGroup:2022pth}.

The decays into dineutrinos, FCNCs mediated by $b \to s \nu \bar\nu$ transitions have not been observed yet, as the final state involves invisibles
which make the search for decays with branching ratios of order $10^{-5}$ or below experimentally challenging. On the other hand,
dineutrino modes have been identified as theoretically very clean probes of the SM for some time, e.g.,~\cite{Grossman:1995gt,Buchalla:1995vs,Buchalla:2000sk,Bartsch:2009qp,Buras:2014fpa}.
Interestingly, the Belle II collaboration recently evidenced for the first time the decay $B^+ \to K^+ \nu \bar \nu$ \cite{talk-Glasov,Belle-II:2023esi}
\begin{align} \label{eq:belleII}
\mathcal{B}(B^+ \to K^+ \nu \bar \nu)=(2.3 \pm 0.7) \cdot 10^{-5} \,, 
\end{align}
in excess of the previously best upper 90 \% CL limit by Belle \cite{Belle:2017oht} 
\begin{align}
\mathcal{B}(B^+ \to K^+ \nu \bar \nu)  < 1.9  \cdot 10^{-5}\,,
\end{align}
and the SM value\footnote{The tree-level background process  $B^+ \to \tau^+ (\to K^+  \bar\nu)\,\nu$ \cite{Kamenik:2009kc} gives a branching ratio of $\sim 5 \cdot 10^{-7}$ \cite{Bause:2021cna}, about ten percent of the FCNC one in the SM \eqref{eq:SMK},
and is negligible at this level of experimental accuracy.}
\begin{align} \label{eq:SMK}
\mathcal{B}(B^+ \to K^+ \nu \bar \nu)_\text{SM}=(4.29 \pm 0.13 \pm \delta_{g}) \cdot 10^{-6} \, .
\end{align}
Here, we updated our previous prediction \cite{Bause:2021cna} by employing the hadronic $B \to K$ transition form factor $f_+$ from a combined fit~\cite{Grunwald:2023nli} of the latest lattice QCD results from the HPQCD collaboration~\cite{Parrott:2022rgu} with results from 
light cone sum rules \cite{Gubernari:2018wyi}. The value \eqref{eq:SMK} is more precise but consistent with previous evaluations~\cite{Bause:2021cna},
also \cite{Becirevic:2023aov,Buras:2022wpw}. We only spell out the form factor uncertainites, as they constitute the main
single source of uncertainty and, what is more important for  the subsequent analysis, the others are from global input such as Cabibbo-Kobayashi-Maskawa (CKM)-elements and 
perturbative Wilson 
coefficients, subsumed in $\delta_g \simeq 0.2$ when added in quadrature, 
which essentially drop out from the test of lepton flavor universality  \cite{Bause:2021cna}.

Taken at face value, the new result \eqref{eq:belleII}, exceeds the SM value by a factor of 3 to 7, and constitutes a deviation around $2.7 \,\sigma$.
The result \cite{talk-Glasov,Belle-II:2023esi} is based on a combination of an inclusive tag analysis using $404\,\text{fb}^{-1}$ yielding an enhanced branching ratio 
$\mathcal{B}(B^+ \to K^+ \nu \bar \nu)_\text{incl}=(2.7 \pm 0.5 (\stat)  \pm 0.5 (\sys)) \cdot 10^{-5}$, with a deviation of $\sim 3.3 \,\sigma$ from the SM,
and one with a hadronic tag, $362\, \text{fb}^{-1}$ of data, with larger uncertainties and consistent with the SM,  
$\mathcal{B}(B^+ \to K^+ \nu \bar \nu)_\text{had}=(1.1^{+0.9}_{-0.8}(\stat) \, \text{}^{+0.8}_{-0.5} (\sys) )\cdot 10^{-5}$.
The enhancement \eqref{eq:belleII} hence relies on the inclusive tag. 
In the future, significantly improved data can be expected from Belle~II ~\cite{Belle-II:2018jsg} to shed light on this anomaly.

The aim of this paper is to explore the implications of an enhanced $\mathcal{B}(B^+ \to K^+ \nu \bar \nu)$ model-independently and 
apply the lepton universality test proposed in Ref.~\cite{Bause:2021cna} to the Belle~II data.

\section{Model-independent interpretation\label{sec:MIA}}

We employ the usual weak effective Hamiltonian for  $b\to s \nu \nu$ transitions 
\begin{equation}
\mathcal{H}_{\text{eff}}=-\frac{4\,G_F}{\sqrt{2}}\frac{\alpha}{4\pi}\sum_{\nu,\nu'} [  C^{\nu \nu'}_L\mathcal{O}^{\nu \nu'}_L + C^{\nu \nu'}_R\mathcal{O}^{\nu \nu'}_R] +\text{h.c.},
\end{equation}
where $C^{\nu \nu'}_{L,R},\,\mathcal{O}^{\nu \nu'}_{L,R}$ denote lepton-specific Wilson coefficients and dimension-six  operators, respectively, renormalized at the  scale $\mu\sim m_b$. 
Furthermore, $G_F$ and $\alpha$ denote the Fermi's constant 
and the finestructure constant, respectively.
The semileptonic  four-fermion operators read
\begin{equation}
\begin{split}
\mathcal{O}^{\nu \nu'}_L&=(\bar{s}\gamma^\mu P_L b)(\bar{\nu}\gamma_\mu  P_L\nu'), \\
\quad \mathcal{O}^{\nu \nu'}_R&=(\bar{s}\gamma^\mu P_R b)(\bar{\nu}\gamma_\mu  P_L \nu'),
\end{split}
\end{equation}
with chiral projectors $P_{L,R}=(1\mp \gamma_5)/2$.
Within the SM, the (lepton universal) Wilson coefficients are given by $C^{\nu \nu'}_{\text{SM}\,L}\,=\,\lambda_t\,X_\text{SM}\,\delta_{\nu \nu'}$, with $X_\text{SM}=-12.64 \pm 0.15$~\cite{Brod:2021hsj} 
and $\lambda_t=V_{tb}V^*_{ts}=-0.0398\pm0.0008$~\cite{ParticleDataGroup:2022pth}.
Further, right-handed currents  $C^{\nu \nu'}_{\text{SM} \, R}$ are suppressed by $m_s/m_b$, and negligible.

We parameterize dineutrino branching ratios in terms of Wilson coefficients and SM input such as hadronic form factors, prefactors from $\mathcal{H}_\text{eff}$ and meson masses, 
see \cite{Bause:2021cna} for details, as
\begin{align} &\mathcal{B}(B^+ \to K^+ \nu \bar \nu) =A^{BK}_+\, x^+\, , \nonumber  \\
&\mathcal{B}(B^0 \to K^{*\,0} \nu \bar \nu) =A^{BK^*}_+\,x^+ + A^{BK^*}_-\,x^- \, , \nonumber  \\
x^\pm  &=\sum_{\nu, \nu'} | C^{\nu \nu'}_{L} \pm C^{\nu \nu'}_{R}   |^2 \,  , \label{eq:xplus}
\end{align}
where $C^{\nu \nu'}_L=C^{\nu \nu'}_{\text{SM} \, L} + C^{\nu \nu'}_{\text{NP} \, L}$ and $C^{\nu \nu'}_R=C^{\nu \nu'}_{\text{NP} \, R}$.
Numerically, we find $A^{BK}_+ =(5.66 \pm 0.17) \cdot 10^{-6}$ employing a recent fit of the associated form factor~\cite{Grunwald:2023nli}.
Moreover, we use $A^{BK^*}_- = (8.88 \pm 1.08) \cdot 10^{-6}$ and $A^{BK^*}_+= (2.00 \pm 0.29) \cdot 10^{-6}$ from Ref.~\cite{Bause:2021cna}.

Due to parity-conservation of the strong interaction the branching ratio of $B \to K \nu \bar \nu$  depends only on $x^+$, i.e.,  $A^{BK}_-=0$.
On the other hand, the decay into strange vectors is predominantly sensitive to the $x^-$ combination,  $A^{BK*}_-  \gg A^{BK*}_+ $ \cite{Hiller:2013cza}.
A combination of both $K$ and $K^*$ final states hence allows to resolve left- and right-handed new physics.
It is also clear that in the absence of right-handed currents $x^-=x^+$ the decays become fully correlated, and
conversely, their correlation is only broken by  the right-handed currents.

We first analyze the impact of  the new measurement of $\mathcal{B}(B^+ \to K^+ \nu \bar \nu)$ alone.
Comparing the data  \eqref{eq:belleII} to the SM \eqref{eq:SMK} one extracts allowed regions of  the Wilson coefficients.
We emphasize that the interpretation depends on the lepton flavor structure.
Assuming lepton flavor universality, the sum in $x^+$ \eqref{eq:xplus} collapses to  a factor three times the universal diagonal entry,
$x^+=3\, |\lambda_t\,X_\text{SM} + C_\text{univ}|^2$. Here, we used for the new physics contribution $C_\text{univ}=C^{\nu \nu}_{\text{NP} \, L}+ C^{\nu \nu}_{\text{NP} \, R} $ (no sum over $\nu$), and obtain  
\begin{align}
C_\text{univ} \in [0.5, 0.8]  \,, ~ C_\text{univ} \in [-1.8, -1.5] \,,  
\end{align}
where the first region corresponds to the one with the SM-sign 
while the second one interferes destructively and hence requires a larger NP effect.

New physics couplings with more general lepton flavor patterns are larger.
In case of universality violation but conserved lepton flavor,
$x^+=2\, |\lambda_t\,X_\text{SM}|^2 +  |\lambda_t\,X_\text{SM} + C_{\nu}|^2$, where  $C_{\nu}= C^{\nu \nu}_{\text{NP} \, L}+ C^{\nu \nu}_{\text{NP} \, R} $ (no sum over $\nu$), denotes a single NP contribution coupling to a
species with flavor $\nu$.
We find the ranges
\begin{align}
C_\nu \in [1.0, 1.7]\,, ~C_\nu \in [-2.7, -2.0] \,.
\end{align}

In the presence of  purely lepton flavor violating couplings  there is no interference with the SM contribution,
$x^+=3\,|\lambda_t\,X_{\text{SM}}|^2 +  |C_{\nu \nu'}|^2$, where  $C_{\nu \nu'}= C^{\nu \nu'}_{\text{NP} \, L}+ C^{\nu \nu'}_{\text{NP} \, R}$ ($\nu, \nu'$ fixed), 
refers to a single NP contribution coupling to species with flavors $\nu$ and $\nu'$.
This yields the ranges 
\begin{align}
C_{\nu\nu^\prime} \in [-2.1, -1.4] \,,  ~ C_{\nu\nu^\prime} \in [1.4, 2.1]  \,.
\end{align}

A combination of both $B \to K^{(*)} \nu \bar \nu$ modes allows to probe $C^{\nu \nu}_{\text{NP}\,L}$ and $C^{\nu \nu}_{\text{NP}\,R}$ individually.  
For the  strange vector the best 90 \% CL upper limit is obtained by Belle~\cite{Belle:2017oht}
\begin{align}  \label{eq:BelleIIK8}
\mathcal{B}(B^0 \to K^{*\,0} \nu \bar \nu)  < 1.8  \cdot 10^{-5} \, . 
\end{align}
It is about a factor of two away from the SM value \cite{Bause:2021cna} 
\begin{align}  \label{eq:SMKst}
\mathcal{B}(B^0 \to K^{*\,0} \nu \bar \nu)_\text{SM}  = ( 8.24  \pm  0.99)   \cdot 10^{-6} \, . 
\end{align}

\begin{figure}
    \centering
    \includegraphics[width=0.49\textwidth]{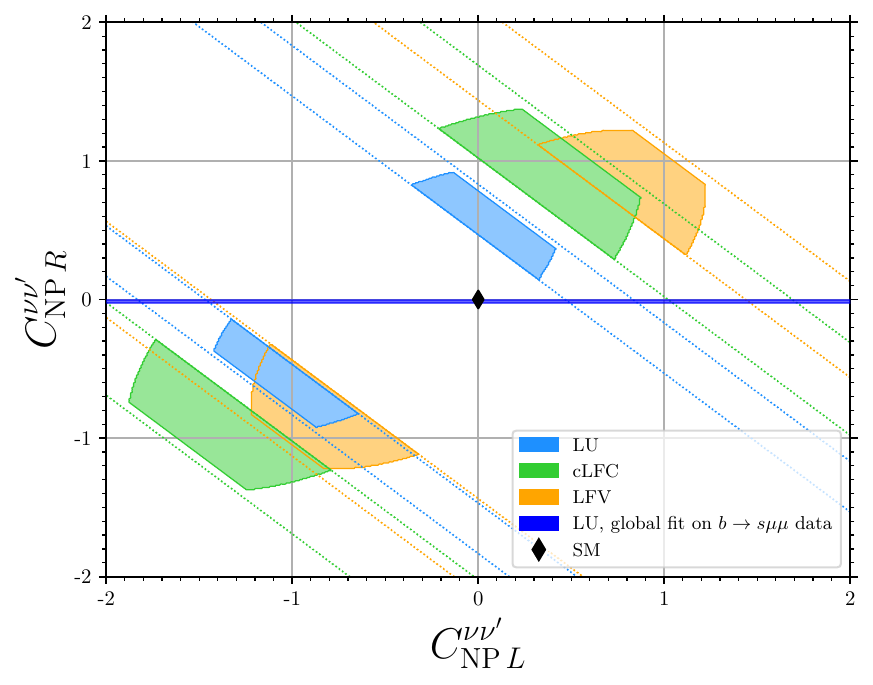}
    \caption{Global constraints on $C^{\nu \nu'}_{\text{NP}\,L}$ and $C^{\nu \nu'}_{\text{NP}\,R}$ assuming LU (blue), cLFC (green), and LFV (orange) from Eq.~\eqref{eq:xpm}. 
    The SM prediction is represented by the black diamond. The dashed lines illustrate the constraints at $1\sigma$ from Eq.~\eqref{eq:belleII}, see the first inequality in Eq.~\eqref{eq:xpm}. 
    In addition, the allowed $1\sigma$-region of $C^{\nu \nu}_{\text{NP}\,R}$ is displayed (darker blue), see Eq.~\eqref{eq:CR_dimuon_fit}, 
    assuming lepton universality and using results from a global fit performed with dimuon data.      
    }
    \label{fig:kappa_plot}
\end{figure}

Using the available information on the $B^+ \to K^+  \nu \bar \nu$  and $B^0 \to K^{*\,0} \nu \bar \nu$ branching ratios, we get the following ranges for $x^\pm$:
\begin{align}\label{eq:xpm}
2.8 \leq x^+ \leq 5.0~, \quad x^- + 0.2\,x^+ \leq 2.0~.
\end{align}
The first inequality is determined by \eqref{eq:belleII}, while the second one by \eqref{eq:BelleIIK8}. 
These limits can be visualized in the $C^{\nu \nu'}_{\text{NP}\,L}$-$C^{\nu \nu'}_{\text{NP}\,R}$ plane by assuming LU, cLFC, and LFV, considering a single lepton flavor contribution in the last two. 
This is illustrated in Fig.~\ref{fig:kappa_plot}. 
One observes that the SM (black diamond) is inconsistent with the data, and that order one and similar values of 
both quark chiralities are required\footnote{  There is a small region allowed for a single cLFC coupling (blue) consistent with $C_{\text{NP}\,L}^{\nu \nu}=0$ with 
$C_{\text{NP}\,R}^{\nu \nu}\in [1.0, 1.3] $.},
\begin{align}\label{eq:WCsim}
C^{\nu \nu'}_{\text{NP}\,R} \sim C^{\nu \nu'}_{\text{NP}\,L} \sim  1 \, ,
\end{align}
which can have either sign.
Assuming tree-level, unsuppressed mediators $ \sim \mathcal{O}_{L,R}/\Lambda^2$, 
this implies a scale $\Lambda$ of new physics to be roughly of 
\begin{equation} \label{eq:M}
\Lambda \sim (5 - 10) \,  \text{TeV} \,,
\end{equation}
which can be probed at the HL-LHC~\cite{ATLAS:2022hsp,CMS:2022cju}.
We also show in Fig.~\ref{fig:kappa_plot} the allowed $1\sigma$-region (blue horizontal line) of $C^{\nu \nu}_{\text{NP}\,R}$ using results from a global fit performed with dimuon data~\eqref{eq:CR_dimuon_fit}, see Ref.~\cite{Bause:2021cna} for details. 
This region gives the strongest constraint on the right-handed  couplings assuming lepton universality. 
We learn that the other lepton universal solutions (light blue areas) are excluded. 

\section{Testing universality}

The full power of the combined analysis of $B \to K^{(*)} \nu \bar \nu$ modes
unfolds once $SU(2)_L$-symmetry is involved, as this enables a combination of the dineutrino modes with processes with charged (left-handed) leptons.
Explicit proposals for charm, beauty and top sectors have been put forward in Ref.~\cite{Bause:2020auq},
and an analysis targeted at $B$-decays in~\cite{Bause:2021cna}.
Interestingly, the method allows to perform novel tests of lepton flavor universality using processes into dineutrino final states,
which are not flavor-tagged and reconstructed as missing energy.

We give here the basic steps of the analysis, and then present the results, harvesting the new data \eqref{eq:belleII},  
together with other improvements over the original study \cite{Bause:2021cna}, where further details can be seen.

The starting point is the correlation through the Wilson coefficients, \eqref{eq:xplus}:
The one-to-one correlation via $C^{\nu \nu}_{\text{NP}\,L}$ between the modes with  pseudo-scalar and vector final state is
broken by the right-handed currents, $C^{\nu \nu}_{\text{NP}\,R}$. We constrain the latter with limits from global $b \to s \gamma, s \mu \mu$ fits~\cite{Bause:2021cna,Bause:2022rrs}.
The  relevant ingredient here is the operator\footnote{At the technical level, this is an operator within the standard model effective theory (SMEFT),
that holds for new physics at a scale $\Lambda$ sufficiently above the  weak scale $v=(\sqrt{2}\,G_F)^{-1/2} \simeq 246\,$GeV.}
$O^{23ij}_{\ell d }=\bar D_2 \gamma_\mu D_3 \bar L_i \gamma^\mu L_j$, where $L$ denotes $SU(2)_L$-doublet-leptons with flavor indices $i,j$ and
$D_{2,3}$ denote down-type $SU(2)_L$-singlet quarks of second and third generation, aka strange and beauty.
$O^{23ij}_{\ell d}$ induces both $O_R^{\nu \nu^\prime}$, 
and right-handed $b \to s \ell_j^+ \ell_i^-$ transitions.
Due to the flavor-blind reconstruction of dineutrino modes  
the lepton mixing matrix cancels out in the amplitude squared due to unitarity \cite{Bause:2020auq}.
We neglect contributions from $Z$-penguin operators which are constrained by electroweak observables, or mixing \cite{Efrati:2015eaa,Brivio:2019ius}.
In our analysis we employ the limits on dimuons, $\ell_{i,j}=\mu$, which are the strongest. 
Imposing this limit for the other lepton flavors and neglecting lepton flavor violation, 
we obtain predictions based on lepton universality.

To see how this works algorithmically, 
we obtain from \eqref{eq:xplus} an analytical formula \cite{Bause:2021cna} 
\begin{align} \label{eq:luregion}
&\mathcal{B}(B^0 \to K^{*\,0} \nu \bar \nu)
    _{\text{LU}}\,=\,\frac{A_+^{BK^*}}{A_+^{BK}}\,\mathcal{B}(B\to K\,\nu\bar\nu)_{\text{LU}}\\
    &\,+\,3\,A_{-}^{BK^*}\,\left(\sqrt{ \frac{\mathcal{B}(B\to K\,\nu\bar\nu)_{\text{LU}}}{3\, A_+^{BK} }}\mp 2\, C^{\nu \nu}_{\text{NP}\,R} \right)^2~,\nonumber
\end{align}
where the subscript `LU' indicates that in this expression lepton universality is understood.
For $C^{\nu \nu}_{\text{NP}\,R} \lesssim \mathcal{O}(1)$ and SM-like $\mathcal{B}(B^+\to K^+\,\nu\bar\nu)$ or larger, Eq.~\eqref{eq:luregion} is well-approximated by
\begin{align} \label{eq:luregionapprox}
\mathcal{B}(B^0 \to K^{*\,0} \nu \bar \nu)_{\text{LU}}  \simeq \frac{A_+^{BK^*} + A_-^{BK^*} }{A_+^{BK}}  \mathcal{B}(B\to K \nu\bar\nu)_{\text{LU}}\,, 
\end{align}
which, of course, is exact if $C^{\nu \nu}_{\text{NP}\,R}=0$. Numerically,
$ (A_+^{BK^*} + A_-^{BK^*})/A_+^{BK} = 1.92 \pm 0.21$.
It becomes apparent that form factor uncertainties entering the $A_\pm$-parameters are the dominant ones.
From a global fit using updated data and the same approach as Ref.~\cite{Bause:2021cna}, 
we obtain using dimuon data 
\begin{align}\label{eq:CR_dimuon_fit}
  C^{\nu \nu}_{\text{NP}\,R}=\lambda_t\,(0.38\pm 0.28) \,.  
\end{align}

\begin{figure}[t]
\centering{
\includegraphics[height=0.37\textwidth]{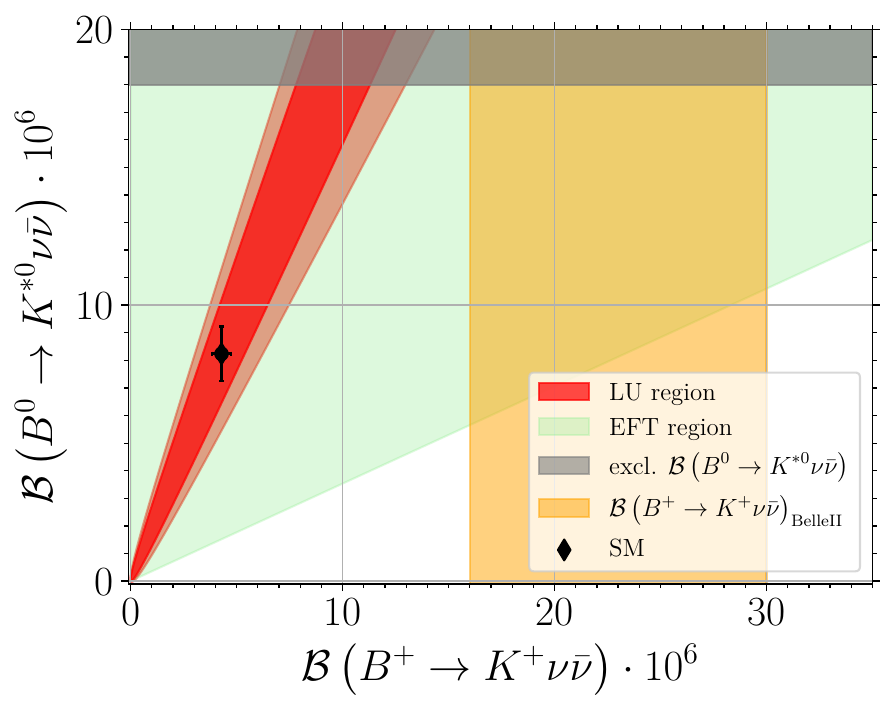}
}
\caption{
Branching ratios of $B \to K^{*\,0} \nu \bar \nu$ and $B^+ \to K^+ \nu \bar \nu$ in the SM (black cross), where for graphical reasons the uncertainty in $\mathcal{B}\left(B^+ \to K^+ \nu \bar \nu\right)$
has been inflated to $10\%$ to the value in \eqref{eq:SMK}.
In addition, the Belle~II data
\eqref{eq:belleII} (vertical orange band) and the 90 \% CL excluded region \eqref{eq:BelleIIK8} (horizontal gray area) are shown. 
The red (dark red) area \eqref{eq:luregion} denotes the $1\sigma$ ($2\sigma$) region that is consistent with lepton flavor universality. 
The green area is the one consistent with \eqref{eq:xplus} and \eqref{eq:boundary}.
}
\label{fig:corr}
\end{figure}

Fig.~\ref{fig:corr} illustrates the central result of our works. 
The correlation between $B^0 \to K^{*\,0} \nu \bar \nu$ and $B^+ \to K^+ \nu \bar \nu$ consistent with lepton universality \eqref{eq:luregion}  (red area),
has no overlap with the Belle~II data \eqref{eq:belleII} (vertical orange band) once the  90 \% CL excluded region \eqref{eq:BelleIIK8} (horizontal gray area) is respected. 
Taken at face value, this means non-universality is back.
Compared to \cite{Bause:2021cna}, the reach of universality conservation (red area) is significantly more narrow,
as a result of the smaller uncertainty in the $B \to K$ form factor $f_+$.
This holds generically for enhanced branching ratios with 
\begin{align} \label{eq:LUV}
\mathcal{B}(B^+ \to K^+ \nu \bar \nu)_{\text{LUV}} > 1.2 \cdot 10^{-5} \, (1.3 \cdot 10^{-5}) \, , 
\end{align}
corresponding to  $1 \sigma$ ($2 \sigma$).
It follows also that the room for lepton universality conservation gets tighter with a decreasing upper limit on, 
or a corresponding future measurement of $\mathcal{B}(B \to K^*\nu \bar\nu)$.
The green area is the one accessible with the EFT parametrization \eqref{eq:xplus}, with boundary
\begin{align} \nonumber
\mathcal{B}(B^0 \to K^{*\,0} \nu \bar \nu)  &\geq A^{BK^*}_+x^+ =\frac{A^{BK^*}_+}{A^{BK}_+}  \mathcal{B}(B^+ \to K^+ \nu \bar \nu) \, ,  \\
& \approx  0.35 \,  \mathcal{B}(B^+ \to K^+ \nu \bar \nu) \, , \label{eq:boundary}
\end{align}
where we have used $x^\pm \geq 0$.
The intersection of the orange band with the green area 
yields a lower limit, 
\begin{align} \label{eq:EFT}
\mathcal{B}(B^0 \to K^{*\,0} \nu \bar \nu)_\text{EFT}   \gtrsim  6\cdot 10^{-6} \, . 
\end{align}

We learn that the correlations matter and also that the smaller details on the SM values for the dineutrinos branching ratios, such as  uncertainties in CKM-input regarding
exclusive or inclusive determinations of $\lambda_t$, which cancel at  leading order \eqref{eq:luregionapprox},
do not play a role in the main conclusions.
The method testing universality works irrespective of this, however the main uncertainties in the $A$-parameters, i.e.~form factors dominate the width of the red area, 
hence govern the room reachable with universality conserving new physics.

\begin{figure}[t]
\centering{
\includegraphics[height=0.3\textwidth]{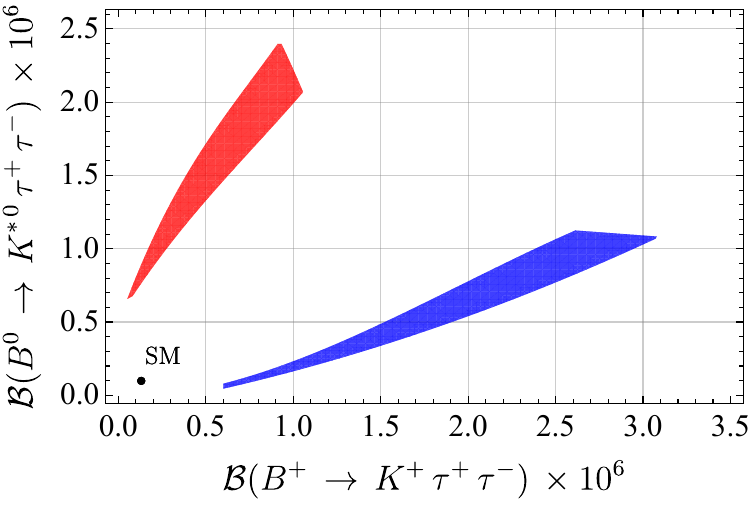}
}
\caption{
Branching ratios of $B \to K^{*\,0} \tau^+\tau^- $ and $B^+ \to K^+  \tau^+\tau^- $ in the SM (black dot)
and  for cLFC heavy NP in taus via $SU(2)_L$  consistent the Belle~II data
\eqref{eq:belleII}  for $C_R>0$ (red) and $C_R <0$ (blue).
The width of the bands and the SM dot reflect theory uncertainties (form factors, CKM matrix elements, and the top mass).  Branching ratios are given for $q^2>15\,\text{GeV}^2$.}
\label{fig:taucorr}
\end{figure}

We also give further $SU(2)_L$-based implications. 
We start with rare decays into taus.
Eq.~\eqref{eq:belleII} implies a large contribution in $b_R\to s_R\tau^+\tau^-$, $C^{\nu \nu'}_{\text{NP}\,R}\sim 1$ for left-handed taus. Considering this NP effect
in  the decays $B_s\to\tau^+\tau^-$, and $B \to X\tau^+\tau^-$ with $X=K,K^*,\phi$, we find
branching ratios as large as 
\begin{align} 
\label{eq:tautau1}
   \mathcal{B}(B_s^0\to\tau^+\tau^-) &\lesssim  1.7 \cdot 10^{-5} \, ,
   \end{align}
   and 
   \begin{align}
     \mathcal{B}(B^+\to K^+ \tau^+\tau^-) & \lesssim 3.1 \cdot 10^{-6}\,, \nonumber \\\label{eq:tautau2} 
       \mathcal{B}(B^0\to K^{*\,0} \tau^+\tau^-) &\lesssim  2.4 \cdot 10^{-6}\,, \\
         \mathcal{B}(B_s^0 \to \phi \,\tau^+\tau^-)& \lesssim 2.2 \cdot 10^{-6} \, ,  \nonumber
\end{align}
where here and in the following in  the semi-tauonic branching ratios  a  cut $q^2>15\,\text{GeV}^2$ in the dilepton invariant mass squared is understood.
We used $C_9^{\prime \, \tau\tau}=-C_{10}^{\prime \, \tau \tau}=\frac{C^{\nu \nu}_{\text{NP}\,R}}{2 \,\lambda_t}$ defined  in the hamiltonian
for dileptons  
$\mathcal{H}_{\text{eff}}=-\frac{4\,G_F}{\sqrt{2}}\frac{\alpha}{4\pi} \lambda_t  (C_9^{\prime \, \tau \tau} \bar s_R \gamma_\mu b_R \bar \tau \gamma^\mu \tau+
C_{10}^{\prime \, \tau \tau} \bar s_R \gamma_\mu b_R \bar \tau \gamma^\mu \gamma_5\tau)$.
Since $|C_{9,10}^{\prime \, \tau \tau}| > |C_{9,10}^{\text{SM}}|$  destructive interference between new physics and the SM contributions can arise  to some extend.
However,
due to the  different sensitivity
to right-handed currents in  decays into vector hadrons, which is similar to the one of  purely leptonic decays, and 
pseudoscalar final hadrons \cite{Hiller:2013cza}, see also  the discussion after (\ref{eq:xplus}),  at least one of them is enhanced, as  demonstrated  in  Fig.~\ref{fig:taucorr}.
This encourages the experimental study  of more than one type of tauonic mode.
Belle~II projections for 5 ab$^{-1}$ (50 ab$^{-1}$) are $\mathcal{B}(B_s^0\to\tau^+\tau^-)\leq 8.1(-)\cdot 10^{-5}$ and $\mathcal{B}(B^+\to K^+\tau^+\tau^-)\leq 6.5(2.0)\cdot 10^{-5}$~\cite{Belle-II:2018jsg}.
Furthermore, at the FCC-ee collider one expects about 1000 $B^0 \to K^{*\,0} \tau^+\tau^-$ events, about a hundred more than at Belle II~\cite{FCC:2018byv}.
Note, $\mathcal{B}(B_s^0\to \tau^+\tau^- )_{\text{SM}}=(6.93\pm 0.30)\cdot10^{-7}$, 
$ \mathcal{B}(B^+\to K^+ \tau^+\tau^- )_{\text{SM}}=(1.17\pm 0.14)\cdot10^{-7}$, $\mathcal{B}(B^0\to K^{*\,0}\, \tau^+\tau^-)_{\text{SM}}= (0.87\pm 0.11)\cdot10^{-7}$,  and $\mathcal{B}(B_s^0\to \phi\, \tau^+\tau^-)_{\text{SM}}=(0.81\pm0.10)\cdot10^{-7}$~\cite{Bause:2021cna},
recalling that we used a kinetic cut on the ditau mass squared $q^2>15\,\text{GeV}^2$.
Eqs.~\eqref{eq:tautau1}, \eqref{eq:tautau2} motivate nearer term searches for  $B_s^0 \to \tau^+\tau^-$, while the semitauonic modes may be sufficiently enhanced by new physics to be seen at Belle II;
they can be studied in detail at the FCC-ee.

Let us briefly consider further NP contributions to the tau modes that are apriori unrelated to the dineutrino modes.
Left-  and right-handed taus interfere in the semileptonic modes \eqref{eq:tautau2} only at  order $(m_\tau/m_B)^2$ or via electromagnetic $b \to s$ dipole operators \cite{Hiller:2003js}, both of which are subleading effects. Therefore, 
no large cancellations between the
different types of tau couplings can occur.
In addition, again, due to the complementarity  between decays into vector and pseudoscalar final hadrons \cite{Hiller:2013cza}
 possible cancellations  of the $b_R s_R\tau_L\tau_L$-contributions with left-handed quark FCNCs   cannot affect both types of decays, so at least one of them should be enhanced.

As $b \to s$  FCNCs into dineutrinos (charged dileptons) 
are $SU(2)_L$-related to $t \to c$ FCNCs with charged dileptons (dineutrinos), 
the NP hint from Belle~II point towards significant new physics effects in $c t \ell \ell$ processes~\cite{Bause:2020auq}.

\section{Light new physics}

The universality test relies on the assumption that new physics is heavy, that is, confined to the Wilson coefficients.
If new physics is light, for instance, sterile neutrinos, or dark fermions, further operators arise, and contribute to
the signal in $B \to K^{(*)} +$ missing energy.
Additional observables and correlations need to be invoked to test this possibility.
Besides invalidating the EFT-bound \eqref{eq:EFT},
light new physics can induce decays of $B_s^0 \to \inv$.

In the presence of scalar (S) and pseudoscalar (P) operators,  
\begin{equation}
\begin{split}
\mathcal{O}^{\nu \nu'}_{S(P)}&=(\bar{s}  P_R b)(\bar{\nu} (\gamma_5) \nu'), \\
\quad \mathcal{O}^{\nu \nu' \, '}_{S(P)}&=(\bar{s} P_L b)(\bar{\nu} (\gamma_5) \nu'),
\end{split}
\end{equation}
as in models with additional  light right-handed neutrinos, the modes $B_s^0\to \nu\bar\nu$ and $B \to K^{(*)}\nu\bar\nu$ become strongly correlated~\cite{Bause:2021cna}
\begin{align}
    \frac{\mathcal{B}(B^0\to K^{*\,0}\nu\bar\nu)_{S,P}}{\mathcal{B}(B^+\to K^+\nu\bar\nu)_{S,P}}&\approx 113\,\left(\frac{y^-}{y^+}\right)~,\label{eq:corr1}\\
    \frac{\mathcal{B}(B_s^0\to \nu\bar\nu)_{S,P}}{\mathcal{B}(B^+\to K^+\nu\bar\nu)_{S,P}}&\approx 79\,\left(\frac{y^-}{y^+}\right)~,\label{eq:corr2}
\end{align}
with the combination of Wilson coefficients
\begin{align}\label{eq:yD}
\begin{split}
    y^\pm=
    \sum_{\nu,\nu^\prime}\bigg(
    \vert \mathcal{C}_S^{\nu\nu^\prime}\pm\mathcal{C}_{S}^{\nu\nu^\prime} \vert^2
    +\vert \mathcal{C}_P^{\nu\nu^\prime}\pm\mathcal{C}_{P}^{\nu\nu^\prime} \vert^2\bigg)~.
\end{split}
\end{align}
Combining \eqref{eq:corr1} and \eqref{eq:corr2}, we obtain
\begin{align}\label{eq:Bsnunu}
    \mathcal{B}(B_s^0\to \nu\bar\nu)_{S,P}&\approx 0.7\,\mathcal{B}(B^0\to K^{*\,0}\nu\bar\nu)_{S,P}\nonumber\\
    &\,<\,1.3\cdot 10^{-5}~, 
\end{align}
where Eq.~\eqref{eq:BelleIIK8} has been employed. Eq.~\eqref{eq:Bsnunu} gives the maximal branching ratio allowed by the current data when including S and P operators.
Presently, no dedicated experimental search for
 $B_s^0\to\inv$ 
exists
\footnote{ Recently an upper limit $\mathcal{B}(B_s^0 \to \inv)< 5.9 \cdot 10^{-4}$  at 90 \% CL  has been obtained  using a recast of LEP-data  \cite{Alonso-Alvarez:2023mgc},
which is weaker than (\ref{eq:Bsnunu}).}, however, the projected sensitivities of Belle~II could help to shed light in this regard, 
$\mathcal{B}(B_s^0\to \nu\bar\nu)< 9.7 \,(1.1)\cdot 10^{-5}$ for 0.12 (0.5) ab$^{-1}$~\cite{Belle-II:2018jsg}.
Note that a possible contribution from Majorana neutrinos is much smaller, ${\cal{B}}(B_s^0 \to \nu \nu) \lesssim 1.2 \cdot 10^{-9}$~\cite{Bortolato:2020bgy}.

\section{Heavy flavorful mediators}

We briefly comment on the challenges for heavy mediator models addressing the $b \to s \nu \bar\nu$ branching ratios.

Since the ratios of muon to electron branching fractions in $b \to s$ transitions, $R_{K,K^*}$ \cite{Hiller:2003js}, are consistent with lepton universality \cite{LHCb:2022qnv},
requisite universality violation to address the data could arise from tau-flavors, or lepton flavor violating ones.
Let us discuss leptoquarks interpretations, 
which are flavorful mediators of new physics and suitable for flavor and collider searches~\cite{Dorsner:2016wpm}.
Leptoquark coupling to a single lepton-species such as tau-flavor can be engineered with flavor symmetries~\cite{deMedeirosVarzielas:2015yxm,Hiller:2016kry}.
Representations that induce tree-level $C_L$ are the scalar singlet $S_1$, the scalar triplet $S_3$ and the vector triplet $V_3$~\cite{Hiller:2016kry}.
Conversely, the doublet scalar $\tilde S_2$ and the doublet vector $V_2$ induce $C_R$ only.
As couplings to both quark chiralities need to be present at similar size, see Fig.~\ref{fig:kappa_plot}, 
this requires at least two different leptoquark representations,
one for $C_L$ and one for $C_R$. 
Models that feed only into one type of coupling cannot address the new Belle~II result in full, for example \cite{Becirevic:2022tsj,deGiorgi:2022vup,Bordone:2017lsy,Asadi:2023ucx,Browder:2021hbl},
as well as those with loop-level contributions to  $b \to s \nu \bar \nu$ transitions \cite{Fuentes-Martin:2020hvc}.

However, in the above lists
there is no representation for $C_L$ that does not also feed into the charged current $b \to c \tau \nu$, subject to constraints.
So in order to achieve a viable model, couplings need to be tuned 
and possibly further degrees of freedom, that is yet another representation or another coupling  as some representations have more than one,
need to be invoked for a complete picture.  This could imply a BSM sector as complex as with three leptoquarks.

To illustrate this we present a leptoquark scenario that induces new physics in $C_L$ with tau-specific $S_1(\bar 3,1,1/3)$ and in $C_R$ with $\tilde S_2(3,2,1/6)$, noting that this model is incomplete and
needs further ingredients, a concrete  example of which is given below, to evade the $b \to c \tau \nu$ constraints.  Here we also spelled out the representations  under the SM group $SU(3)_C \times SU(2)_L \times U(1)_Y$, and denote the leptoquark mass by $M_R$ where the representation $R=S_1, \tilde S_2,..$ is indicated by the subscript.

Using \cite{Dorsner:2016wpm}, we obtain for $C_L$  from $S_1$
\begin{align} \label{eq:CLS1}
C_L^{S_1} = 
\frac{\pi}{\alpha} \frac{v^2}{M_{S_1}^2} y_1^{b \tau} y_1^{s \tau \, *} \, .
\end{align}
Similarly, for $\tilde S_2$, which corresponds to  $\tilde R_2$ in  \cite{Dorsner:2016wpm},
\begin{align}
C_R^{\tilde S_2} = 
- \frac{\pi}{\alpha} \frac{v^2}{M_{\tilde S_2}^2} y_2^{s \tau} y_2^{b \tau \, *} \, . 
\end{align}
Using \eqref{eq:WCsim} and with order one Yukawas $y_1, y_2$ this gives  leptoquark masses around $5 \, \text{TeV}$. 
This is above present LHC limits \cite{CMS:2022zks}.

To complete the $S_1, \tilde S_2$-scenario  one could add a vector singlet $V_1(3,1,2/3)$. This leptoquark representation  does not induce
tree-level $b\to s \nu \bar \nu$ but $b \to c \tau \nu$, and can suppress BSM effects in the latter by
cancelling the $S_1$-induced contribution to the charged-current  mode with 
\begin{align}
 \frac{y_1^{b \nu} y_1^{c \tau \, *} }{2 M_{S_1}^2}  \simeq - \frac{ x_1^{c \nu} x_1^{b \tau \, *} }{M_{V_1}^2}  \, .
\end{align}
We also note that instead of $S_1$ one could have used $S_3(\bar 3,3,1/3)$ for $C_L$, with  Yukawa $y_3$, and 
\begin{align}
C_L ^{S_3}= 
\frac{\pi}{\alpha} \frac{v^2}{M_{S_3}^2} y_3^{b \tau} y_3^{s \tau \, *} \, ,
\end{align}
and yukawa-mass dependence as $S_1$ \eqref{eq:CLS1}, and corresponding tuning with $V_1$,
\begin{align}
 \frac{y_3^{b \nu} y_3^{c \tau \, *} }{2 M_{S_3}^2}  \simeq + \frac{ x_1^{c \nu} x_1^{b \tau \, *} }{M_{V_1}^2}  \, .
\end{align}
Unlike the $S_1$, the triplet  $S_3$
also induces tree-level $\bar s_L b_L \tau^+\tau^-$, hence enhances  the branching ratios  for ditau modes  \eqref{eq:tautau1}, \eqref{eq:tautau2},
which are based on right-handed currents only,  by up to a factor of two.

\section{Conclusions}
The recent measurement of the $B^+ \to K^+ \nu \bar \nu$ branching ratio by Belle~II points to a value enhanced over the SM one.
While this apparently challenges the SM, we show that this challenges lepton flavor universality as well, unless new physics is light.
This conclusion holds generally for any $\mathcal{B}(B^+ \to K^+ \nu \bar \nu)$ in excess of the limit \eqref{eq:LUV},
which includes the experimental finding \eqref{eq:belleII}. 
This is manifest from Fig.~\ref{fig:corr}, 
showing that there is no overlap between the regions allowed by
$B^+ \to K^+ \nu \bar \nu$ (orange band), $B \to K^{*\,0} \nu \bar \nu$ (below the gray band) and intersecting the universality reach (red region).

Since the ratios of muon to electron branching fractions in $b \to s$ transitions, $R_{K,K^*}$ \cite{Hiller:2003js}, are consistent with lepton universality \cite{LHCb:2022qnv}, 
requisite universality violation to address the data could arise from tau-flavors. 
More general explanations could also involve lepton flavor violation, which strengthens the motivation for searches in $B \to K^{(*)} \tau^+\tau^-$, $B_s^0 \to \tau^+\tau^-$, $B \to K^{(*)} \tau \ell$, $\ell=e,\mu$. 
We stress however that the requisite mass scale is quite low \eqref{eq:M}, and therefore explicit models with tree-level mediators (leptoquarks, $Z^\prime$) need
to be carefully balanced to avoid other flavor constraints, notably those from  $b \to c \tau \nu$ data. 
This implies very involved model building with non-minimal particle content. We discuss this for leptoquark models, and
find that more than two different type of couplings, or representations are required. 
The $ b \to s \mu \mu$ anomalies in the angular distributions and reduced semileptonic branching ratios would need additional 
lepton flavor-specific new physics.

The enhanced $B^+ \to K^+ \nu \bar \nu$ branching ratio \eqref{eq:belleII} reported~\cite{talk-Glasov,Belle-II:2023esi} relies on the inclusive tagging analysis.
Further experimental analyses with higher statistics are required to shed light on this anomaly.
In addition, searches for the decay $B_s^0 \to \inv$ should be pursued as limits on its branching ratio are helpful to understand whether new physics is heavy, 
or light degrees of freedom not captured by the weak effective theory contribute to $B \to K\,\inv$.
The new Belle~II result could imply branching ratios $\mathcal{B}(B_s^0\to \nu\bar\nu)$ as large as
$10^{-5}$. 

\bigskip 

{ \bf Note added:}
During the final stages of this work a related article~\cite{Athron:2023hmz} on the Belle~II measurement appeared.

\bigskip

\mysection{Acknowledgements}
We are happy to thank Daniel Litim and Lara Nollen for useful discussions. 
This work was performed at Aspen Center for Physics (GH), which is supported by National Science Foundation grant PHY-2210452. 
The work of HG was supported by the project ``CPV-Axion'' under the Supporting TAlent in ReSearch@University of Padova (STARS@UNIPD) and the INFN Iniziative Specifica APINE.

\end{document}